\documentclass[11pt,a4paper]{article}
\usepackage[hmargin=25mm,vmargin=22mm]{geometry}
\usepackage[T1]{fontenc}
\usepackage{graphicx,float}
\usepackage{lmodern,microtype,amsmath,amssymb,booktabs,tabularx,array,xcolor}
\usepackage[colorlinks=true,linkcolor=black,citecolor=black,urlcolor=blue!45!black]{hyperref}
\newcommand{\sect}[1]{\par\addvspace{0.65em}{\large\bfseries #1}\par\nobreak\vspace{0.1em}}
\newcolumntype{Y}{>{\raggedright\arraybackslash}X}
\begin{document}
\begin{center}
{\LARGE\bfseries It Is Not My Code Anymore}\par
\vspace{0.55em}
{\small Augusto Camargo}\par
{\small Institute of Mathematics, Statistics and Computer Science (IME)}\par
{\small University of S\~ao Paulo, S\~ao Paulo, Brazil}\par
{\small 7 September 2026}
\end{center}
{\small\itshape A research note on authorship, responsibility, and evaluation in AI-assisted software production. The discussion uses a hypothetical failure and a selective reading of the literature; it reports no new empirical results.}

\begin{figure}[H]
\centering
\includegraphics[width=\linewidth]{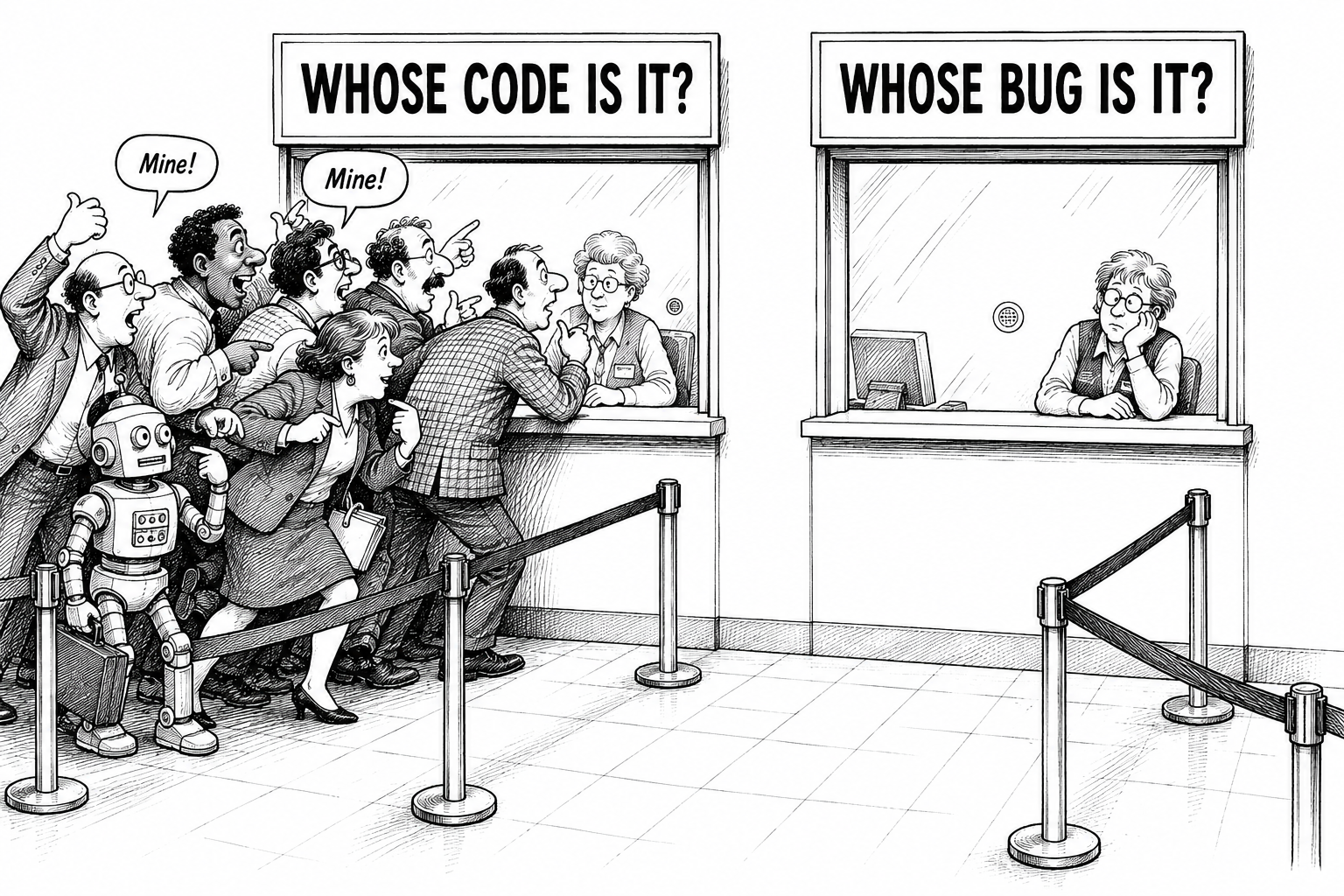}
\caption{Whose code, whose bug? A satirical contrast between claiming authorship and accepting responsibility after failure, not an empirical finding of the cited studies.}
\label{fig:whose-code-whose-bug}
\end{figure}

Suppose a large language model (LLM) accidentally introduces \texttt{1 / 0} while implementing a feature requested by a developer. The developer reads the diff without paying close attention and approves the change. Evaluating the expression raises an unhandled exception. During enrollment, the program reaches the expression and fails. A student cannot enroll.

\emph{Who wrote that expression?}

The developer may have specified the objective, selected the model, guided its revisions, and accepted its output. In this example, however, the model produced the expression. Reviewing it did not change who produced it.

Responsibility for the failure is a separate question. We still need to examine the review, the release process, the harness that coordinates model and tool use, and the organization operating the service. Knowing who produced the expression does not tell us who should have caught the defect or who must restore the service.

I consider what a developer means by \emph{my code} when a model produces it, and how shared ownership relates to responsibility after failure. I then ask whether the task needs a separate program at all. Throughout, the practical requirement is the same: the student must be able to enroll.

\sect{Authorship: who produced the code?}

Calling a vehicle ``my ride'' may mean that I use it to get somewhere. It does not establish that I designed it, built it, or own it. Similarly, ``my code'' can refer to code I wrote, inherited, maintain, or am expected to repair. The possessive alone does not distinguish these relationships. Taking over maintenance does not make someone the author of an existing expression. Generative models make this familiar ambiguity harder to ignore.

Seo, Deldari, and Mentis examine these relationships in \emph{Whose Code Is It?} In a study with 30 participants, increasing AI contribution reduced perceived possession and shifted attribution toward AI. Willingness to accept accountability for production systems remained comparatively stable~\cite{seo}. Draxler and colleagues report a related distinction in text generation: users can feel little ownership of generated text while publicly declaring themselves its authors~\cite{draxler}. Feeling ownership, claiming authorship, and accepting responsibility can be separate judgments.

The human contribution still matters. Naur's account of programming as theory building emphasizes understanding~\cite{naur}; Meske and colleagues describe a shift toward orchestration through probabilistic intent mediation~\cite{meske}. Directing an implementation and producing it are distinguishable contributions. An art director can shape a work without executing every mark. In the opening example, the developer directed the work and the model supplied the expression.

Reviewing a diff may fulfill an assigned duty and support a release decision. It does not make the reviewer the producer of an expression already in the diff. This is the sense in which I use ``It is not my code anymore.'' The phrase does not assign legal ownership to a model.

Review can change whether code is released without changing the code itself. Once the expression runs, the language and execution environment determine its behavior. A probabilistic generator can produce a fully reproducible defect. Calling it a hallucination adds little to the immediate investigation: how did the expression enter the program, and why did the resulting failure reach the student?

\sect{Responsibility after failure}

Tholander and Jonsson describe programming with generative AI as a co-constituted practice in which intentions, authorship, and control develop through interaction~\cite{tholander}. That description is compatible with tracing a particular expression to a particular source. Several contributors can produce a service together even when the model introduced the defective expression.

\emph{Reconstructing who contributed to a program does not, by itself, determine who has which responsibilities after its failure.}

Different production histories can yield the same program, and the recorded history may be incomplete. Even a complete record would leave questions about each participant's duties and control over the process.

In the opening example, an investigation must establish who introduced the expression and which controls should have detected or contained it. It must also identify who owed the student a functioning service and who must restore it. Each question can have a different answer. Saying ``we co-created it'' leaves those answers unspecified.

Tracing the expression further back through model training and training-data production adds to its causal history. That history may help an investigation, but it does not assign the same responsibility to everyone involved. Assigning responsibility requires identifying a relevant contribution, an applicable duty, and evidence connecting the two.

Collective psychological ownership describes a group's relationship to something it regards as its own~\cite{pierce}. After a failure, the team also needs clear duties: who can stop a release, who must question the acceptance evidence, and who investigates, repairs, and explains. Elish's \emph{moral crumple zone} describes the risk of blaming a nearby human who had limited control~\cite{elish}. Santoni de Sio and Mecacci distinguish several responsibility gaps rather than treating responsibility as a single issue~\cite{gaps}. Shared ownership alone does not resolve these gaps.

Would people attribute the code in the same way after a successful delivery and after a failure? Figure~\ref{fig:whose-code-whose-bug} asks this question through satire. An answer would require evidence about behavior after failure, beyond the relationships examined in the programming studies. Using ownership to organize production still leaves duties, controls, and investigation procedures to be specified. Release decisions and incident handling depend on these details. We therefore need to examine how model-generated implementation changes the production process.

\sect{We put the machine in the loop}

People already specified, implemented, reviewed, released, and maintained software using machines. A generative model now joins that process as a producer of implementations. From this perspective, the change is \emph{machine-in-the-loop}. Starting with the model and asking where to insert a human reverses the order in which the process developed.

Preventing the generated defect from reaching the student requires controls across production. Industrial history offers a precedent for reorganizing those controls when work is divided differently. The American Society for Quality (ASQ) links the division of craft work and Taylorist production to changes in inspection~\cite{asq}. Shewhart's control-chart work, dated to 1924 by the National Institute of Standards and Technology (NIST), shifted attention toward process variability~\cite{nist}. Quality did not begin with mechanization, but its organization changed.

For generated software, examining production means considering the model, configuration, context, harness, tools, dependencies, and human interventions together. A model or harness change can alter production even when the requested function stays the same. Reviewing one output provides evidence about that output. Evaluating later outputs requires evidence from repeated production runs.

Collecting and using this evidence requires an assigned role. I would give quality engineering a central role: defining acceptance evidence, questioning the process, tracking failures, and supporting release decisions with explicit criteria. The role needs clear duties and decision authority. Software already has assurance disciplines; the question is how their responsibilities change when implementation is generated.

This work extends beyond reviewing the finished diff. U.S. Food and Drug Administration (FDA) guidance states that inspection and testing alone cannot assure pharmaceutical quality~\cite{fda}. Applied to software production, the principle is to examine both the generating process and its outputs. Statistical evaluation can help when the populations and measurements are justified. A probabilistic generator does not make unrelated programs a homogeneous production batch.

\sect{Evaluating the service and its production}

A process could consistently produce code that passes local checks while leaving the student unable to enroll. Repeatability alone would not establish that the service meets the demand. Acceptance criteria therefore need a reference beyond the producing process.

\textbf{Inside-out} evaluation starts with the producer, the code, and their properties, then extends toward the service. \textbf{Outside-in} evaluation starts with the required outcome and conditions of use, then identifies the evidence needed from each contributing system.

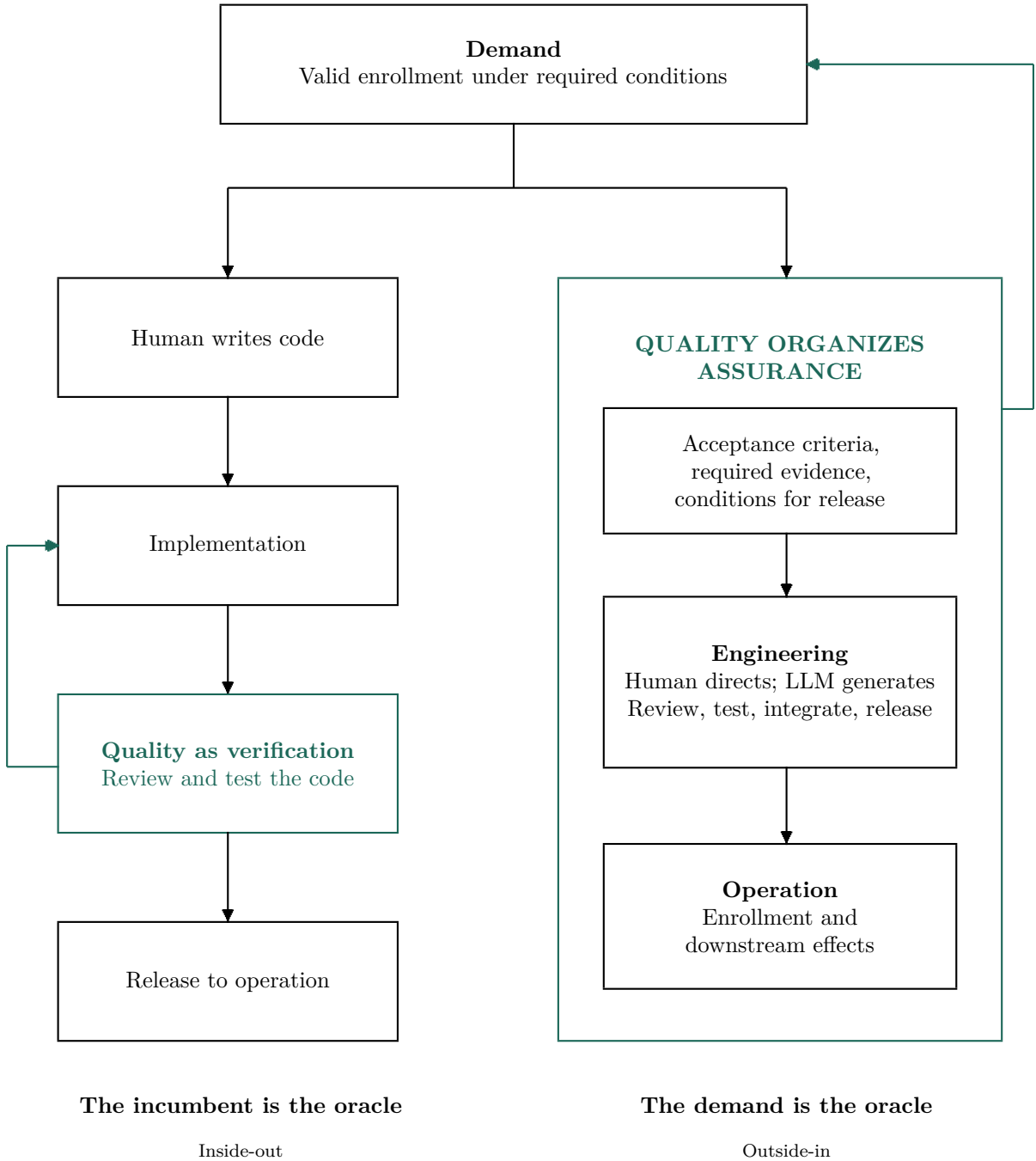
\begin{figure}[p]
\centering
\begingroup
\setlength{\unitlength}{1mm}
\definecolor{qualityscope}{RGB}{27,103,88}
\newcommand{\nodebox}[3]{\framebox(#1,#2){\parbox{#1mm}{\centering\small #3}}}
\newcommand{\downhead}{\multiput(-0.9,1.8)(0.1,-0.2){1}{\rule{1.8mm}{0.25mm}}\put(-0.75,1.5){\rule{1.5mm}{0.35mm}}\put(-0.6,1.2){\rule{1.2mm}{0.35mm}}\put(-0.45,0.9){\rule{0.9mm}{0.35mm}}\put(-0.3,0.6){\rule{0.6mm}{0.35mm}}\put(-0.15,0.3){\rule{0.3mm}{0.35mm}}}
\newcommand{\righthead}{\put(-1.8,-0.9){\rule{0.35mm}{1.8mm}}\put(-1.5,-0.75){\rule{0.35mm}{1.5mm}}\put(-1.2,-0.6){\rule{0.35mm}{1.2mm}}\put(-0.9,-0.45){\rule{0.35mm}{0.9mm}}\put(-0.6,-0.3){\rule{0.35mm}{0.6mm}}\put(-0.3,-0.15){\rule{0.3mm}{0.3mm}}}
\newcommand{\lefthead}{\put(1.5,-0.9){\rule{0.35mm}{1.8mm}}\put(1.2,-0.75){\rule{0.35mm}{1.5mm}}\put(0.9,-0.6){\rule{0.35mm}{1.2mm}}\put(0.6,-0.45){\rule{0.35mm}{0.9mm}}\put(0.3,-0.3){\rule{0.35mm}{0.6mm}}\put(0,-0.15){\rule{0.3mm}{0.3mm}}}
\begin{picture}(160,178)
\thicklines
\put(35,160){\nodebox{90}{18}{\textbf{Demand}\\Valid enrollment under required conditions}}
\put(80,160){\line(0,-1){10}}
\put(36,150){\line(1,0){86}}
\put(36,150){\vector(0,-1){14}}
\put(122,150){\vector(0,-1){14}}
\put(10,118){\nodebox{52}{18}{Human writes code}}
\put(36,118){\vector(0,-1){14}}
\put(10,86){\nodebox{52}{18}{Implementation}}
\put(36,86){\vector(0,-1){14}}
\put(10,51){\color{qualityscope}\nodebox{52}{21}{\textbf{Quality as verification}\\Review and test the code}}
\put(36,51){\vector(0,-1){14}}
\put(10,19){\nodebox{52}{18}{Release to operation}}
\put(10,61){\color{qualityscope}\line(-1,0){8}}
\put(2,61){\color{qualityscope}\line(0,1){34}}
\put(2,95){\color{qualityscope}\vector(1,0){8}}
\put(87,19){\color{qualityscope}\framebox(68,117){}}
\put(90,120){\makebox(62,8){\parbox{62mm}{\centering\small\textcolor{qualityscope}{\textbf{QUALITY ORGANIZES\\ASSURANCE}}}}}
\put(94,97){\nodebox{54}{19}{Acceptance criteria,\\required evidence,\\conditions for release}}
\put(122,97){\vector(0,-1){10}}
\put(94,61){\nodebox{54}{26}{\textbf{Engineering}\\Human directs; LLM generates\\Review, test, integrate, release}}
\put(122,61){\vector(0,-1){12}}
\put(94,27){\nodebox{54}{22}{\textbf{Operation}\\Enrollment and\\downstream effects}}
\put(155,116){\color{qualityscope}\line(1,0){5}}
\put(160,116){\color{qualityscope}\line(0,1){53}}
\put(160,169){\color{qualityscope}\vector(-1,0){35}}
\put(0,6){\makebox(76,6){\small\textbf{The incumbent is the oracle}}}
\put(0,-1){\makebox(76,6){\scriptsize Inside-out}}
\put(84,6){\makebox(76,6){\small\textbf{The demand is the oracle}}}
\put(84,-1){\makebox(76,6){\scriptsize Outside-in}}
\put(36,136){\downhead}
\put(122,136){\downhead}
\put(36,104){\downhead}
\put(36,72){\downhead}
\put(36,37){\downhead}
\put(122,87){\downhead}
\put(122,49){\downhead}
\put(10,95){\color{qualityscope}\righthead}
\put(125,169){\color{qualityscope}\lefthead}
\end{picture}
\endgroup
\caption{Both flows start with the demand. Green return arrows show what guides acceptance: the implementation on the left and the demand on the right. Black arrows show the forward flow. On the right, quality organizes assurance of engineering and operation using demand-derived criteria; the green boundary marks that scope. The comparison concerns the reference for acceptance. It is not an exhaustive history or an automatic consequence of using an LLM. Human-driven implementation can also be demand-centered.}
\label{fig:assurance-scope}
\end{figure}

For enrollment, outside-in evaluation asks whether software production supports the required outcome. Component checks still provide evidence, but the demand determines what that evidence must establish. Figure~\ref{fig:assurance-scope} contrasts quality as a verification stage with a broader role in organizing assurance of engineering and operation, including model-based production.

\emph{The Caf\'e in Amsterdam} asks whether the incumbent or the demand supplies the oracle~\cite{cafe}. Here, the incumbent is a view of software work centered on code. Adding review and oversight around the code preserves that starting point. Enrollment may also depend on academic records, payment, access rights, service recovery, and administrative action. The student's demand crosses these boundaries.

Test-driven development can start before implementation while remaining within a software layer. Tests of individual operations can pass even when the enrollment journey fails. An oracle that checks only the confirmation screen can report success without a persistent enrollment record.

Testing the steel in a car's body provides evidence about a material. A proving ground tests the assembled vehicle under required conditions. Both provide useful evidence, but they evaluate different objects against different criteria. The distinction is independent of whether testing is automated or performed early.

A software proving ground could exercise the enrollment service through user journeys: different legitimate paths, concurrent requests, interruptions, and recovery attempts. Its oracle would check valid enrollment and required downstream effects, not simply responsiveness. A thousand simulated users help only if their behavior represents the relevant conditions. Repeating one synthetic behavior a thousand times does not establish that match. Table~\ref{tab:acceptance-evidence} gives examples of evidence required beyond successful component behavior.

\begin{table}[H]
\centering
\caption{Component evidence and acceptance evidence for the enrollment service.}
\label{tab:acceptance-evidence}
\small\renewcommand{\arraystretch}{1.15}
\begin{tabularx}{\linewidth}{@{}YY@{}}
\toprule
Evidence about a part & Evidence required by the demand \\
\midrule
The generated change passes its tests & The eligible student obtains a valid enrollment \\
The service responds under load & Enrollment completes within the required conditions and time \\
The retry handler executes & Retrying causes no duplicate enrollment or charge \\
The interface reports success & The academic record and required access confirm that success \\
\bottomrule
\end{tabularx}
\end{table}

The acceptance criteria in Table~\ref{tab:acceptance-evidence} need to be justified against the demand. Deriving every expected result from the generated implementation would make the evaluation depend on the system being evaluated.

The same criteria can evaluate a delivery and the process that produced it. Artifact evaluation asks whether this delivery works. Process evaluation asks how consistently a recorded production setup delivers acceptable results. To evaluate the process, produce multiple implementations under recorded configurations and evaluate them against the same externally justified criteria.

Representative operating scenarios and deliberately adversarial scenarios serve different purposes. Failures found by actively searching for problems do not directly estimate failure frequency in ordinary use.

\sect{Does the task need a separate program?}

Evaluating against the demand allows us to compare different ways of providing the service. If acceptance criteria specify what a valid enrollment requires, they need not assume a particular implementation. We can therefore ask whether meeting those criteria requires generating a program specifically for the task.

Three possibilities need to be distinguished. A program can exist while its code is hidden from the user. Code can also be generated dynamically as the task is performed. In both cases, an implementation is still produced. Hassan and colleagues envision intent-centric development with code secondary and hidden by default~\cite{hassan}. Rost explores computing through reflective interaction, including dynamically generated code~\cite{rost}. These changes concern how people interact with code and when it is produced; they do not necessarily remove it.

The third possibility is that an existing trained system performs the task without generating a new program for it. Welsh's vision of trained systems replacing much conventional software motivates this possibility~\cite{welsh}. This does not mean dispensing with software: the trained system still depends on an implementation and operating infrastructure. The question is whether each task needs an additional, separately produced program. Generating machine code instead of source code would still produce a task-specific program.

For enrollment, the distinction is between generating a program to carry out the required operations and having an existing system carry them out without generating that program. Whether the latter can meet the demand is an evaluation question, not an assumption. It would need to satisfy the same acceptance criteria, including a valid enrollment record and the required downstream effects, under its actual operating conditions.

Where this substitution is feasible, there is no newly generated program whose authorship needs to be assigned. Responsibility for providing the service remains, including responsibility for the system's dependencies and failures. The question about ownership of generated code thus gives way to the same practical question that guided the evaluation: does the chosen means of delivery meet the demand?

\sect{Limitations and conclusion}

The division by zero is a hypothetical example, not an incident report. The cited studies do not show that collective ownership collapses after failure, nor do they claim to certify safe delivery. I ask how the relationships they describe inform release decisions and incident handling.

A proving ground covers selected conditions of use and can miss relevant ones. Its scenarios and acceptance criteria need justification against the conditions in which the service is required. Observation after deployment is needed to detect omissions and revise the tests.

In the example, the model produced the expression, a production process accepted it, and the student experienced the failure. Distinguishing these facts lets us ask who must act, what they must do, and what evidence is needed before the next release.

``It is not my code anymore'' changes how I describe my relationship to the implementation. Calling the code ours, or replacing the program with another means of delivery, still leaves the service to be evaluated. The student must be able to enroll.

\end{document}